\documentclass[aps,pre,showpacs,showkeys,twocolumn,groupedaddress]{revtex4}
\usepackage{graphicx}
\usepackage[utf8]{inputenc}
\usepackage[english]{babel}
\usepackage{amsmath}
\usepackage[T1]{fontenc}
\usepackage{amssymb}
\usepackage{units}
\usepackage{float}

\newcommand{\mat}[1]{\MakeUppercase{#1}}

\newcommand{\co}[1]{#1}			

\begin{document}

\title{Spreading in Integrable and Non--integrable Many--Body Systems}
\author{Johannes Freese, Boris Gutkin and Thomas Guhr}
\date{\today}
\affiliation{Fakult\"at f\"ur Physik, Universit\"at Duisburg--Essen,
                   Lotharstra\ss e 1, 47048 Duisburg, Germany}

\begin{abstract}
  We consider a finite, closed and selfbound many--body system in
  which a collective degree of freedom is excited. The redistribution
  of energy and momentum into a finite number of the non--collective
  degrees of freedom is referred to as spreading as opposed to damping
  in open systems. Spreading closely relates to thermalization, but
  while thermalization requires non--integrability, spreading can also
  present in integrable systems.  We identify subtle features which
  determine the onset of spreading in an integrable model and compare
  the result with a non--integrable case.
\end{abstract}


\maketitle

\section{Introduction}
\label{sec1}

Almost sixty years ago, Fermi, Pasta and Ulam~\cite{FPU} presented a
puzzling study of a many--body system which contrary to the intuition
does not thermalize: the phase space is not ergodically filled. They
studied an one--dimensional chain of point particles coupled by
springs with a small nonlinear force.  It was found numerically that
the energy which was put into the lowest Fourier mode almost
completely stayed there, even after very long times, rather than being
distributed over many or all modes. An understanding was achieved in a
continuum limit which leads to a Korteweg--de Vries equation allowing
for soliton solutions.  A recent review can be found in
Ref.~\cite{BI}. Some years ago, Kinoshita, Wenger and Weiss~\cite{KWW}
studied a different but related problem in a real experiment. They
realized a quantum Newton's cradle by letting two one--dimensional
Bose gases collide and oscillate against each other.  Once more,
thermalization did not take place, the two Bose gases kept their
shapes even after many oscillations.  Subsequently this lack of
thermalization was attributed to the integrability of the problem.
This conclusion however is sometimes challenged because the system is
only weakly interacting and can thus be viewed as two sub-systems (the
centers of mass of the two Bose gases) which collide without effect on
their inner structure~\cite{KR}.  Thermalization and related issues
are presently an active field of research, see
Refs.~\cite{quench,cosme} and references therein.  The notion of
thermalization is sometimes restricted to infinite many--body
systems. Here, we use it in a broader sense, including large but
finite many--body systems as, for example, Bose gases and atomic
nuclei.

Thermalization can be accompanied by another phenomenon --
\textit{spreading} of collective motion. It is often observed in
closed (and finite) many--body systems that exhibit collective and
incoherent single--particle motion simultaneously. Energy and momentum
from one distinct --- in the present context collective --- degree of
freedom is redistributed into many other single--particle degrees of
freedom.  This has to be distinguished from \textit{damping} as
occurring in open systems. In the latter case, the energy leaves the
system because of coupling to a large or even infinite number of
external degrees of, \textit{i.e.}, to an external bath as studied in
the generic Caldeira--Leggett model~\cite{Caldeira:1982iu}.

\begin{figure}[h]
\includegraphics[width=0.4\textwidth]{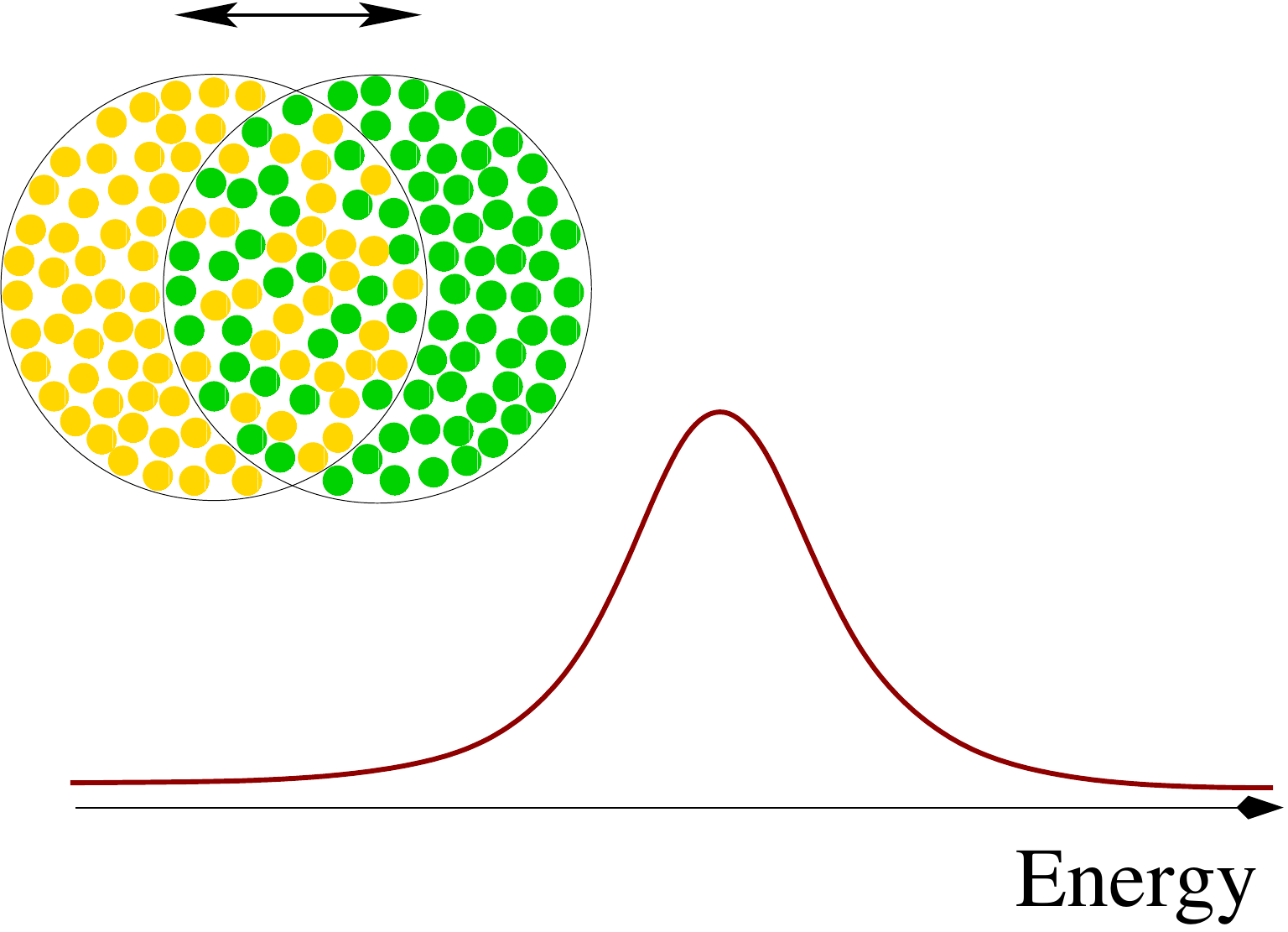}
\label{fig0}
\caption{Schematic sketch of a Giant Dipole Resonance. The
     peak is the envelope of a very large number of individual
     quantum mechanical resonances which cannot be resolved
     in an experiment.}
\end{figure}

Among the numerous examples for spreading of collective motion in
atomic nuclei~\cite{Bohr}, the Giant Dipole Resonance, as schematically
depicted in Fig.~\ref{fig0},
is probably the most well--known one. The
cross section of the electric dipole radiation as well as the spectral
density of the excitations show at a certain energy a huge peak whose
spreading width is orders of magnitudes larger than the mean level
spacing. A simple, somehow semiclassical picture helps to catch the
salient features of this effect: The neutrons may be viewed as
confined to one sphere, the protons to another one. Relative motion of
the nucleons inside these spheres does not take place. The two spheres
move against each other, thereby carrying out a fully collective
motion. This results in an enormous response function.  A bit further
away from the resonance energy, relative motion inside the spheres
sets in which lowers the cross section.  Much further away, all motion
is incoherent of single--particle type, the motion is not collective
anymore, and the resonance has disappeared completely.  Besides the
Giant Dipole Resonance, many other forms of collective motion and the
associated spreading exist in nuclei, see recent examples in
Refs.~\cite{End,End2}. Further examples of collective motion can also
be found in Bose--Einstein condensates~\cite{KWW,But99,Mad00,Mar00}.
Here, we study an effect inspired by the Giant Dipole Resonance in a
simple system of interacting particles.

As opposed to thermalization, chaos is not crucial for the presence or
absence of spreading. Spreading of collective motion might occur even
in integrable systems~\cite{hgg,hggl}. However, as we show in the
present study, the details of this process are very sensitive to the
choice of the system parameters. 
In particular, we demonstrate that in the integrable case  pronounce spreading occurs only for fine tuned interactions between system particles.  
We also study numerically how the onset of chaos
influences both, collective and single--particle dynamics.

The paper is organized as follows. In Sec.~\ref{sec2}, the model is
introduced and the procedure of its numerical solution are discussed.
We present the results for the integrable and non--integrable cases in
Secs.~\ref{sec3} and~\ref{sec4}, respectively. We conclude in
Sec.~\ref{sec5}.

\section{Setup of the Model}
\label{sec2}

We largely use the model  introduced in our  previous
analytical investigations~\cite{hgg,hggl}. Two clouds of interacting particles in one dimension  are coupled to each other. 
In the integrable case, all interactions between particles are harmonic. A forth--order term is  added then   to  explore a weakly non--integrable
regime. The clouds are initially
separated and then released, which does or does not lead to a
spreading of the initial energy and momentum over all degrees of
freedom.  

 As we found it appropriate
to slightly change some conventions of \cite{hgg,hggl} for the numerical study, we
compile all necessary formulas defining the Hamiltonian in
Sec.~\ref{sec21}. We discuss the choice of collective coordinate in
Sec.~\ref{sec22}. The numerical method and the initial conditions are
explained in the Secs.~\ref{sec23} and~\ref{sec24}, respectively.

\subsection{Hamiltonian}
\label{sec21}

The two clouds, labeled $a=1,2$, of $N$ point particles each with
equal masses $m$, move in one dimension.  Their positions and momenta are denoted
$x_i^{(a)}, \ i=1,\ldots,N$ and  $p_i^{(a)}, \
i=1,\ldots,N$, respectively. The total Hamiltonian is
\begin{equation}
H = H_0 + \lambda  H_\textrm{ni} \ .
\label{Hamiltonian}
\end{equation}
The first part of $H$ is the integrable Hamiltonian as discussed in Ref.~\cite{hgg},
\begin{equation}
H_0 = H_{0}^{(1)}+ H_{0}^{(2)} + \kappa H_{0}^{(12)} \ , 
\label{H0}
\end{equation}
  where the  terms
\begin{equation}
H_{0}^{(a)} = \frac{1}{2m}\sum^N_{i=1}\left(p^{(a)}_i\right)^2 +
                                              \sum^N_{i\neq j} V_{ij}\left(x^{(a)}_i-x^{(a)}_j\right)^2 \ , 
\label{eq2}
\end{equation}
model the two harmonic clouds $a=1,2$ which are coupled with the  interaction term
\begin{equation}
H_{0}^{(12)} = \sum^N_{i,j=1}K_{ij} \left(x^{(1)}_i - x^{(2)}_j \right)^2  \ .
\label{eq3}
\end{equation}
As already pointed out, we focus on selfbound
systems.  Here, we find it convenient to ensure this directly by using
the translation invariant differences of the positions in Eqs.~(\ref{eq2},\ref{eq3}). 
The selfboundness is ensured since we choose both the coefficients
$K_{ij}$, $V_{ij}$ from  symmetric  $N\times N$ matrices $K, V$ with positive entries.  In
contrast to Refs.~\cite{hgg,hggl} we introduce the control parameter
$\kappa$ for tuning the overall strength ratio of the interactions
 between the clouds. 
Finally, the second term $\lambda H_\textrm{ni}$  in~\eqref{Hamiltonian}  renders the total
Hamiltonian $H$ non--integrable.

We order the positions and momenta in two $2N$ component vectors
$x=(x^{(1)},x^{(2)})$ and $p=(p^{(1)},p^{(2)})$ with the $N$ component
vectors $x^{(a)}=(x_1^{(a)},\ldots,x_N^{(a)})$ and
$p^{(a)}=(p_1^{(a)},\ldots,p_N^{(a)})$ for $a=1,2$.  Furthermore, it
is helpful to cast the integrable case for $\lambda=0$ into a more
compact form.  Defining the $2N \times 2N$ positive, symmetric
interaction matrix
\begin{eqnarray}
C = \begin{bmatrix}  W & -\kappa K/2 \\
                                     -\kappa K/2 & W
       \end{bmatrix} \ ,
\label{eqC}
\end{eqnarray}
with
\begin{equation}
W_{ij} = \left(2\sum_{l=1}^N V_{il}+\kappa K_{il}\right)\delta_{ij} - V_{ij},
\end{equation}
the potential becomes a standard bilinear form and we arrive at the
expression
\begin{equation}
H_0 = \frac{p^2}{2m} + x^T Cx \ .
\label{H0C}
\end{equation}
To fix the notation, we write down the elementary transformation to
normal modes explicitly.  An orthogonal matrix $U$ diagonalizes the
interaction matrix,
\begin{eqnarray}
C &=& \frac {m}{2} U^T \omega^2 U  \nonumber\\
\omega &=& {\rm diag\,}(\omega_1,\ldots,\omega_{2N}) \ ,
\label{dia}
\end{eqnarray}
where the eigenvalues $m\omega_i^2/2, \ i=1,\ldots,2N$ of $C$ are
non--negative, because the matrices $V$ and $K$ have positive entries.
In the rotated coordinates
\begin{equation}
\xi = U x   \qquad {\rm and} \qquad   \pi = U p
\label{rot}
\end{equation}
the system Hamiltonian decouples into $2N$ non--interacting ones,
\begin{equation}
H_0 = \sum_{i=1}^{2N} \left(\frac{\pi_i^2}{2m} + \frac{1}{2}m\omega_i^2 \xi_i^2\right) \ .
\label{H0Crot}
\end{equation}
The positive quantities $\omega_i$ are of course the system
eigenfrequencies. The coordinates $\xi_i$ and $\pi_i$ are not positions
and momenta of the particles, rather they are weighted linear
combinations and can be viewed as positions and momenta of the
non--interacting composite particles which define the normal modes.
The transformation \eqref{H0Crot} also facilitates an elementary
solution of the equations of motions, allowing for a crucial check of
our numerics later on.

In Ref.~\cite{hggl}, we extended the integrable model \eqref{H0}  by adding a rather general translation invariant term
 which breaks the integrability but preserves the
selfboundness. For our numerical study we make the special choice
$\lambda H_\textrm{ni}$ with a strength parameter $\lambda$ and the
fourth--order potential
\begin{equation}
H_\textrm{ni} = \sum^N_{i,j=1} P_{ij} \left(x^{(1)}_i - x^{(2)}_j \right)^4 \ .
\label{eq4}
\end{equation}
The coefficients $P_{ij}$ are as well taken from a  symmetric
$N\times N$ matrix $P$ with positive entries. This non--integrable interaction preserves
translation invariance and selfboundness.

\subsection{Collective Coordinate}
\label{sec22}

We aim at studying the interplay between collective and incoherent
single--particle motion. Many--body systems show a rich variety of
collective excitations, particularly nuclei provide a zoo of
examples~\cite{Bohr}. Of course the way how the system is probed
determines which collective modes are excited. As we have in mind
excitation in which the two clouds are simply pulled apart and then
released to oscillate against each other, the natural choice for the
collective coordinate in our case is the difference of the centers of
mass in each cloud
\begin{equation}
\Xi = \frac{1}{N} \sum_{i=1}^{N} x_i^{(1)}  - \frac{1}{N} \sum_{i=1}^{N} x_i^{(2)} \ .
\label{eq5}
\end{equation}
Although this definition is fully equivalent to the one we used
previously, we notice that the collective coordinate $X$ in
Ref.~\cite{hggl} differs by a factor, \textit{i.e.}, we have $\Xi =
\sqrt{N/2} X$.  The time evolution $\Xi(t)$ of the collective
coordinate is our most important observable. The larger the typical
amplitudes $|\Xi(t)|$ of the collective motion after some time $t$,
the more of energy and momentum is contained in the oscillation
between the two clouds. The smaller the amplitudes, the more of energy
and momentum is transferred to the incoherent single--particle degrees
of freedom within the clouds.

\subsection{Numerical Solution}
\label{sec23}

For the numerical integration of the equations of motion we found it
efficient to use the Velocity Verlet Method, see
\textit{e.g.}~Ref.~\cite{Hinch}, a standard method in molecular
dynamics.  We employed various well--established techniques and tests
to implement it in an optimal way for our system. In particular, we
carefully checked that the energy is conserved even for very long
times beyond those we were interested in.  As already mentioned in
Sec.~\ref{sec21}, we compared the exact solution of the integrable
model with a direct numerical integration which turned out highly
useful to eliminate even subtle errors. To further test the results of
our simulation we validated that our numerical simulation remains
stable under time reversal transformation.

\subsection{Initial Conditions}
\label{sec24}

\begin{figure}
\begin{center}
	\includegraphics[scale=0.6]{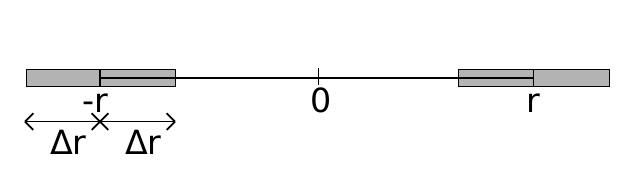}
	\caption{The initial positions of the particles are randomly
          chosen within the shaded intervals.}
\label{fig1}
\end{center}
\end{figure}

The initial conditions are chosen such that the two clouds are
separated and at rest at time $t=0$. The initial particle positions
$x_{i0}^{(1)}=x_i^{(1)}(0), \ x_{i0}^{(2)}=x_{i}^{(2)}(0), \
i=1,\ldots,N$ are taken from two uniform random distributions around
 some points, $r$ and $-r$, within the intervals $[r-\Delta r,
r+\Delta r]$ and $[(-r)-\Delta r, (-r)+\Delta r]$, respectively. This
is illustrated in Fig.~\ref{fig1}. 
In all our investigations, we used mirrored initial conditions,
\textit{i.e.}, the symmetry $x_{i0}^{(2)}=-x_{i0}^{(1)}, \
i=1,\ldots,N$.  The initial particle momenta $p_{i0}^{(1)}, \
p_{i0}^{(2)}, \ i=1,\ldots,N$ are always set to zero when we
investigate the integrable case in Sec.~\ref{sec3}. Only in the
non--integrable case to be discussed in Sec.~\ref{sec4}, we work with
non--zero initial momenta, preserving the mirror symmetry.

\section{Integrable Case}
\label{sec3}

We begin by demonstrating in Sec.~\ref{sec31} the presence and absence
of spreading in the integrable case. In Sec.~\ref{sec32}, we give a first
explanation by looking at eigenfrequencies and normal modes. We 
study the influence of the standard deviations of the distributions for
the interaction matrix elements and of the particle number in 
Secs.~\ref{sec33} and~\ref{sec34}, respectively.

\subsection{Presence and Absence of Spreading}
\label{sec31}

\begin{figure}
\begin{center}
  \includegraphics[scale=1]{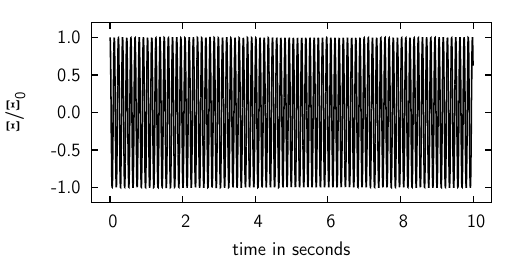}
  \includegraphics[scale=1]{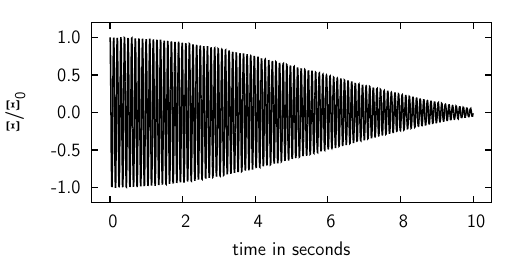}
  \caption{The ratio of collective coordinate to its initial value for $\kappa=0.75$ (top) and
    $\kappa=1.0$ (bottom) versus time from 0 to 10 seconds.}
\label{fig2}
\end{center}
\end{figure}

\begin{figure}
\begin{center}
	\includegraphics[scale=1]{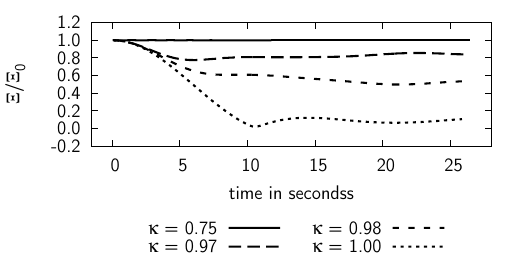}
	\caption{Envelope of the collective coordinate for four values
          of $\kappa$ versus time from 0 to 10 seconds.}
\label{fig3}
\end{center}
\end{figure}

We choose the entries of the interaction matrices $W$ and $K$ from
independent Gaussian distributions with means $\langle W\rangle,
\langle K\rangle$ and standard deviations $\sigma_W, \sigma_K$,
respectively.   The  mass of each particle is always set to be  1~kg. Other parameters for  numerical simulations are given
in Tab.~\ref{results:parameter_for_integrable}.
\begin{table}[H]
    \centering
    \begin{tabular}{c|c|c|c|c}
		$N$ & $T$ & $\Delta t$ & $\langle \mat{W} \rangle$ & $\sigma_W$ \\ 
        \hline
		$500$ & $\unit[30]{s}$ & $\unit[0.001]{s}$ & $\unit[4]{J/m^2}$ & $\unit[0.63]{J/m^2}$ \\
    \end{tabular} 
    \begin{tabular}{c|c|c|c}
		$\langle \mat{K} \rangle$ & $\sigma_K$ & $|\co{r}|$ & $\Delta \co{r}$ \\ 
        \hline
	        $\unit[2]{J/m^2}$ & $\unit[0.48]{J/m^2}$ & $\unit[2]{m}$ & $\unit[0.2]{m}$ \\
    \end{tabular} 
    \caption{Parameter set for the integrable case.}
    \label{results:parameter_for_integrable}
\end{table}
The parameter $\kappa$ defining the strength ratio of the interaction
within and between the clouds is still to be fixed. In
Fig.~\ref{fig2}, the collective coordinate is shown for two different values
of $\kappa$. Here and in all other figures, the collective coordinates $\Xi(t)$
are normalized to their initial values $\Xi_0$. As seen in Fig.~\ref{fig2},
the results differ drastically -- for $\kappa=0.75$, the energy stays in
the collective oscillation, while it is spread over the other degrees
of freedom for $\kappa=1.0$. It is instructive to plot the envelopes of
the collective coordinate, this is done in Fig.~\ref{fig3} for four
values of $\kappa$. Surprisingly, the
transition from weak to almost complete spreading happens within a
relatively small interval of $0.03$ or so in the parameter $\kappa$.
If $\kappa$ is increased beyond $\kappa=1.0$, the spreading becomes
weaker again.

\subsection{Eigenfrequencies and Normal Modes}
\label{sec32}

\begin{figure}
\begin{center}
  \includegraphics[scale=1]{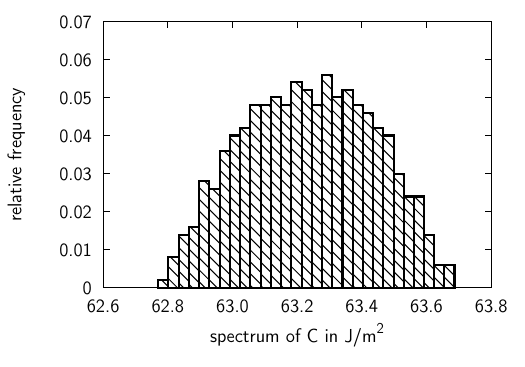}
  \includegraphics[scale=1]{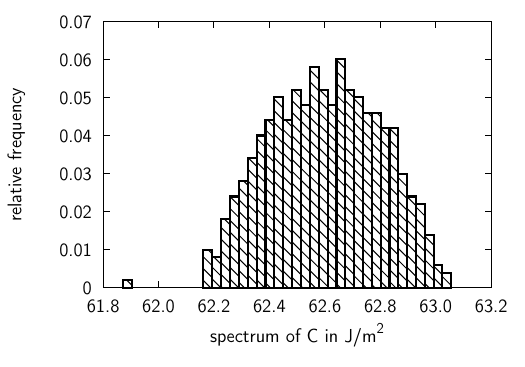}
  \caption{Spectrum of system eigenfrequencies for $\kappa=1$ (top)
    and $\kappa=0.98$ (bottom).}
\label{fig4}
\end{center}
\end{figure}

The above results can be explained by the structure of the interaction
matrix. As the eigenfrequencies \eqref{dia} completely determine the
interaction, we display their spectra for two values of $\kappa$ in 
Fig.~\ref{fig4}.
We notice that half of the eigenfrequencies are not excited, which is
due to the mirrored initial conditions.  This symmetry causes the
projection to the antisymmetric eigenvectors to vanish.  If using
non--mirrored initial conditions, any energy stored in the
corresponding degrees of freedom is completely decoupled from the
collective coordinate. Hence, it does not have an effect on the
dynamics of the collective coordinate. Accordingly, these frequencies  are not shown in
Fig.~\ref{fig4}. 

 Although both spectra for $\kappa=1$ and $\kappa=0.98$ show broad bulks of
non--degenerate eigenfrequencies, they are distinctly different. The
one for $\kappa=1$ has an isolated eigenfrequency left of the
bulk. It is found to belong to the eigenvector that is close  to the
``collective'' vector $e$ , \textit{i.e.}, the one where the modulus of all entries is
equal, but the signs differ for the two clouds.  

\begin{figure}
\begin{center}
  \includegraphics[scale=1]{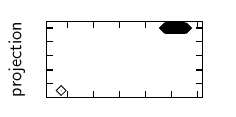}
  \includegraphics[scale=1]{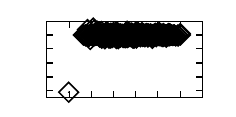}
  \includegraphics[scale=1]{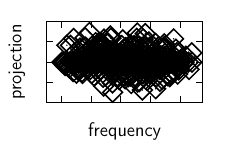}
  \includegraphics[scale=1]{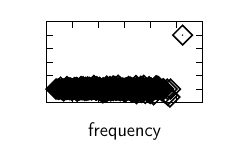}
  \caption{Excitation of the normal modes for $\kappa=0.75, 0.97$
    (top) and $\kappa=1.0, 1.02$ (bottom) versus the
    eigenfrequencies.}
\label{fig5}
\end{center}
\end{figure}

A general equation for the time evolution of the collective coordinate  $\Xi(t)$ can be easily written down 
by using eigenvectors $c_i$ $i=1,\dots ,N$ of the interaction matrix $C$:
\begin{equation}
\Xi(t)=\sum_{i=1}^N a_i \cos (\omega_i t), \label{newequation1}
\end{equation}  
where  $a_i=(e,c_i)(c_i,x_0)$ and the two scalar products correspond to the projection of the eigenmode $c_i$ 
on the ``collective'' vector and  the initial state, respectively.
Note that, since the initial state $x_0$ is aways taken to be close to (a scaled) vector $e$ (see fig.~\ref{fig1}), 
 both scalar products  are similarly distributed. So 
the
crucial information on the dynamics of $\Xi(t)$   can be extracted from  the distribution of the initial excitations  
$\xi_{i0}=(c_i,x_0)$. 
In Fig.~\ref{fig5}, we display the excitation of
the normal modes for four different values of $\kappa$. As there is a
one--to--one correspondence between the amplitude $\xi_{i0}$ and the
eigenfrequency $\omega_i$, we show the initial amplitudes $\xi_{i0}$
versus the eignfrequencies $\omega_i$ to directly visualize which
eigenmode is excited. Comparing with Fig.~\ref{fig3}, we see that
spreading is obviously absent if only the collective excitation is
excited, whereas a broader excitation of other eigenmodes leads to
spreading. The truly amazing observation is the subtlety of this
process which takes place within a 3\% change of the parameter
$\kappa$ and is thus due to minor changes in the structure of the
interaction.  We also conclude that the isolation of the
eigenfrequency corresponding to the collective coordinate is essential
to prevent spreading. This is seen in Fig.~\ref{fig5} for
$\kappa=1.02$. The crucial eigenfrequency now shows up on the right
hand side of the spectral bulk. As a function of $\kappa$ it wandered
through the bulk, it is isolated again, leading to a suppression of
spreading.

\begin{figure}
\begin{center}
	\includegraphics[scale=0.6]{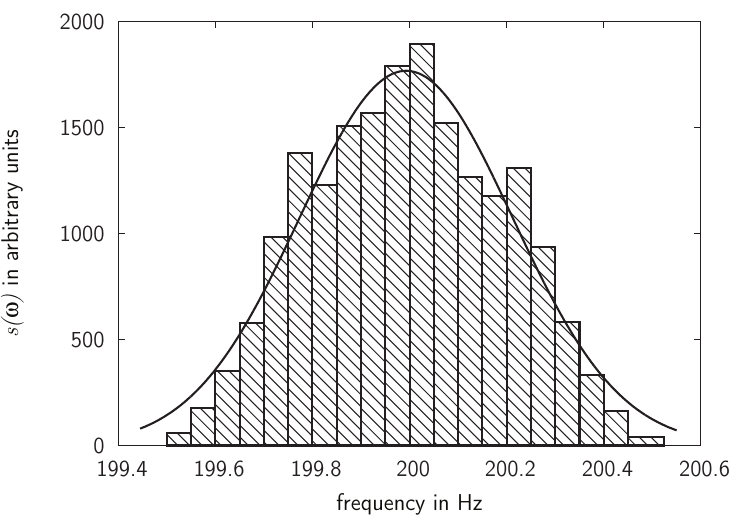}
	\caption{The figure shows a Gaussian fit of $s(w)$ for $N=5000$ particles. Other parameters of the model are the same as in Tab.~\ref{results:table:parameter_for_variance_test}. }
\label{New_Fig1}
\end{center}
\end{figure}

If no isolated eigenvalues are present, the dynamical evolution of the collective coordinate can 
be estimated by eq.~\ref{newequation1} with a smooth
approximation to the spectral density $s(\omega')=\sum_{i=1}^N a_i \delta(\omega'-\omega_i) $, as a function of
the continuous variable $\omega'$,
\begin{eqnarray}
\Xi(t)
          &=& \int\limits_{-\infty}^{+\infty} d\omega'\cos(\omega't) 
                                  \sum_{i=1}^N a_i \delta(\omega'-\omega_i) \nonumber\\
           &\simeq&   \int\limits_{-\infty}^{+\infty} d\omega'\cos(\omega't) s(\omega') \ .                     
\label{collapp1}
\end{eqnarray}
This amounts to a Fourier transform of the spectral density. As we
choose the interaction matrix elements from Gaussian distributions, it
is not too surprising that a Gaussian with mean $\mu$ and width
$\gamma$ approximates the spectral density well, see fig.~\ref{New_Fig1},
\begin{eqnarray}
\Xi(t) &\sim&   \int\limits_{-\infty}^{+\infty} d\omega'\cos(\omega't) 
                                           \exp\left(-\frac{(\omega'-\mu)^2}{2\gamma^2}\right) \nonumber\\
          &\sim&   \exp\left(-\frac{\gamma^2t^2}{2}\right) \cos(\mu t) \ ,                     
\label{collapp2}
\end{eqnarray}
leading to the  Gaussian decay of the collective coordinate with the oscillation period $2\pi/\mu$ and  the decay
time given by $1/\gamma$.  The standard deviations $\sigma_W$ and
$\sigma_K$ in turn determine the width $\gamma$ of the spectral
density.

\subsection{Modifying the Standard Deviations of the Interactions}
\label{sec33}

\begin{figure}
\begin{center}
	\includegraphics[scale=0.55]{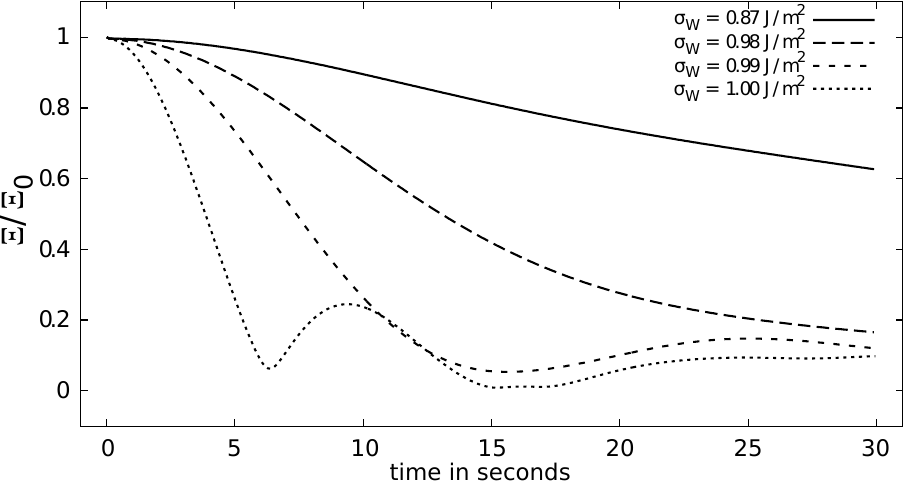}
	\caption{Envelopes of the collective coordinates versus time
          for four different values of $\sigma_W$.}
\label{fig6}
\end{center}
\end{figure}

We further investigate the remarkable sensitivity of spreading to
slight variations of the interaction matrices $W$ and $K$. In view of
its high dimension, we refrain from trying to explore the space of the
interaction parameters systematically. We rather focus on some
examples in which we modify the standard deviations $\sigma_W$ and
$\sigma_K$ of the probability distributions for the elements of the
interaction matrices $W$ and $K$. The parameters for these numerical
simulations are given in
Tab.~\ref{results:table:parameter_for_variance_test}.
\begin{table}[H]
    \centering
    \begin{tabular}{c|c|c|c|c|c|c|c}
        $N$ & $T$ & $\Delta t$ & $\langle \mat{W} \rangle$ & $\langle \mat{K} \rangle$ 
                             & $\co{r}$ & $\Delta \co{r}$ \\
        \hline
	$500$ & $\unit[30]{s}$ & $\unit[0.001]{s}$ & $\unit[2]{J/m^2}$ & $\unit[1]{J/m^2}$ 
                             & $\unit[2]{m}$ & $\unit[0.2]{m}$ \\
    \end{tabular}
    \caption{Parameter set for the test of the standard deviation dependence.}
    \label{results:table:parameter_for_variance_test}
\end{table}
We notice that the relative strength parameter is now fixed to
$\kappa=2$. Furthermore, we found it convenient to keep the
standard deviation of the interaction between the clouds constant,
$\sigma_K=0.1$, and  only to vary the standard deviation $\sigma_W$ of
the interaction within the clouds. In Fig.~\ref{fig6}, the envelopes
of the collective
coordinates are shown for four values of $\sigma_W$. The smaller the
standard deviation $\sigma_W$, the narrower the distribution of the
eigenfrequencies, resulting in the system being less likely to show
spreading.  Again, it is surprising that even relatively small changes
in $\sigma_W$ have a strong impact. This behavior remains the same, if
$\sigma_K$ is changed and $\sigma_W$ is held fixed.

\subsection{Dependence on the Particle Number}
\label{sec34}

\begin{figure}
\begin{center}
	\includegraphics[scale=1]{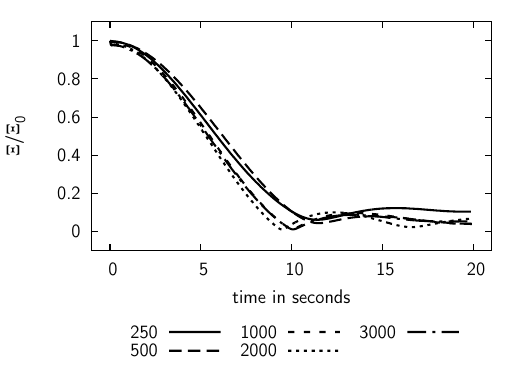}
	\label{fig8}
\end{center}
\caption{Envelopes of the collective coordinates versus time
          for five different particle numbers $N$.}
\end{figure}

\begin{figure}
\begin{center}
	\includegraphics[scale=1]{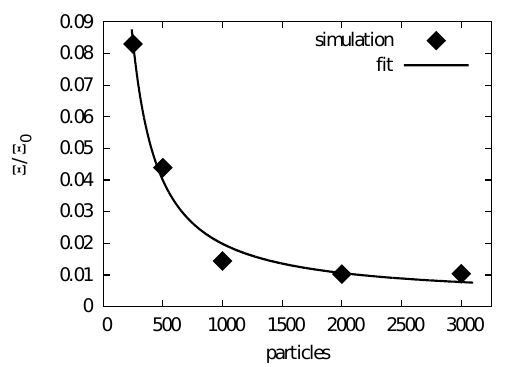}
	\caption{Normalized amplitude of the collective coordinate at
          the bottleneck position versus particle number $N$.}
\label{fig9}
\end{center}
\end{figure}

One is tempted to expect, based on observations in statistical
mechanics, that the number of particle itself is important for the
decay of the collective coordinate: the larger the number of degrees
of freedom, the larger the recurrence times and the more effective
ought to be the process of thermalization. Accordingly, a large
particle number should make spreading more efficient, and the position
of the bottleneck should decrease with the number of particles. In
Fig.~\ref{fig8}, however,
we see a different behavior. The parameters for these numerical
simulations are listed in Tab.~\ref{results:n_dependence}. 
\begin{table}[H]
    \centering
    \begin{tabular}{c|c|c|c|c|c}
        $T$ & $\Delta t$ & $\langle \mat{W} \rangle$ & $\langle \mat{K} \rangle$ 
                     & $\co{r}$ & $\Delta \co{r}$ \\
        \hline
	$\unit[3]{s}$ & $\unit[0.001]{s}$ & $\unit[4]{J/m^2}$ & $\unit[2]{J/m^2}$ 
            & $\unit[2]{m}$ & $\unit[0.2]{m}$ \\
    \end{tabular}
	\caption{Parameter set for the particle number dependence}
    \label{results:n_dependence}
\end{table}
The number of particles $N$ strongly affects the local timescales,
resulting in ever faster oscillations when $N$ grows. This is so
because the total mass of the system increases linearly with $N$,
while the number of interactions for a given particle with other
particles goes with $N^2$. Hence, the oscillations periods decrease.
As argued above, the global timescale for the spreading depends on the
standard deviations of the interactions. For the present choice of
parameters, these means that the decay time and thus position of the
bottleneck is roughly the same for all particle numbers.  However,
another expectation from statistical mechanics manifests itself in
these simulations. As Fig.~\ref{fig9} illustrates, the bottleneck
becomes
sharper with increasing particle number.

\section{Non--Integrable Case}
\label{sec4}

In Sec.~\ref{sec41}, we investigate and compare non--integrable
perturbations of different strengths, before we take a closer look at
the trajectories in the phase space by slightly varying the initial
conditions in Sec.~\ref{sec42}.

\subsection{Perturbations of Different Strengths}
\label{sec41}

\begin{figure}
\begin{center}
  \includegraphics[scale=0.9]{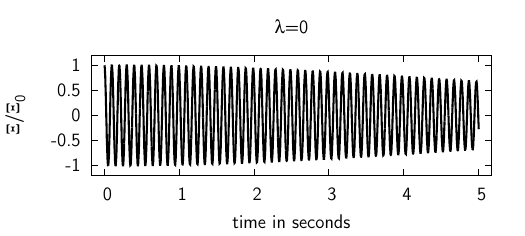}
	\includegraphics[scale=0.9]{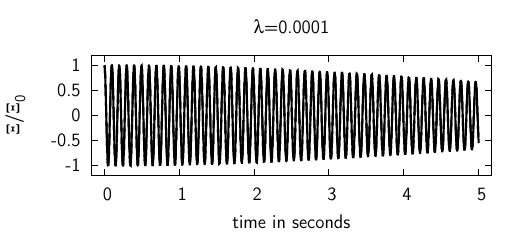}
  \includegraphics[scale=0.9]{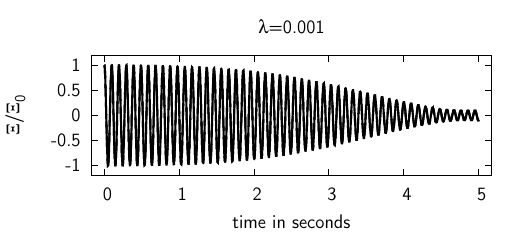}
  \includegraphics[scale=0.9]{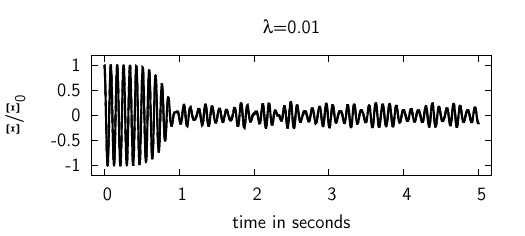}
	\caption{Collective coordinates for different strength $\lambda$ of the non--integrable perturbation.}
  \label{fig10}
\end{center}
\end{figure}

We use the full Hamiltonian~\eqref{Hamiltonian} with the
non--integrable part \eqref{eq4}.  The interaction matrices $W$ and
$K$ for the integrable harmonic interaction, the parameter $\kappa$,
as well as the interaction matrix $P$ for the non--integrable
perturbation are kept fixed in all simulations to be presented here.
The parameters are given in
Tab.~\ref{results:table:parameter_for_kappa_test}.
\begin{table}[H]
    \centering
    \begin{tabular}{c|c|c|c|c|c|c}
        $N$ & $T$ & $\Delta t$ & $\langle w \rangle$ & $v_w$ & $\langle k \rangle$ & $v_k$ \\
        \hline
        $500$ & $\unit[20]{s}$ & $\unit[0.0005]{s}$ & $\unit[2]{J/m^2}$ & 
                      $\unit[0.2]{J/m^2}$ & $\unit[1]{J/m^2}$ & $\unit[0.1]{J/m^2}$ \\
    \end{tabular}
    \begin{tabular}{c|c|c|c|c}
	$\langle P \rangle$ & $v_P$ & $\co{r}$ & $\Delta \co{r}$ & $\Delta v$ \\
	\hline
	$\unit[1.5]{J/m^2}$ & $\unit[0.15]{J/m^2}$ & $\unit[2]{m}$ & $\unit[0.2]{m}$ & $\unit[0.0001]m/s$ \\
	\end{tabular}
    \caption{Parameter set for the investigation of the perturbation influence.}
    \label{results:table:parameter_for_kappa_test}
\end{table}
Only the parameter $\lambda$ is varied to investigate the impact of
different perturbation strengths. The results of the simulations 
are displayed in Fig.~\ref{fig10}. For reference, the integrable
\begin{figure}
\begin{center}
	\includegraphics[scale=1.0]{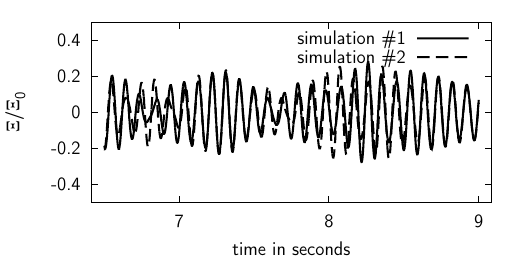}
\label{fig12}
\end{center}
\caption{Collective coordinate for $\lambda=10^{-2}$ at larger
          times. The ordinate is different from the one in
          Fig.~\ref{fig10} where the short--time behavior is shown.}
\end{figure}
case corresponding to $\lambda=0$ is also shown. Obviously, the
non--integrability helps the spreading considerably: for the strongest
perturbation, the bottleneck is reached much quicker than in all
previous simulations. However, as we demonstrated in Sec.~\ref{sec3},
non--integrability is not a necessary condition for spreading.

\subsection{Slight Variations of Initial Conditions}
\label{sec42}

High sensitivity of the trajectories to slight changes in the initial
conditions is the prime signature of classical
chaos~\cite{gutzwiller_chaos}.  The proper measure is the Lyapunov
exponent.  In this spirit, we now measure the distance between 
 system trajectories which differ only slightly in the initial
conditions. We look at the collective coordinate as well as on some
single--particle trajectories.  When varying the initial conditions,
we  ensure that the total energy of the system remains
unchanged to carry out a comparison on equal footing. To this end, we
realized the changes in the initial conditions by modifying the
particle momenta instead of the particle positions.  We recall that in
the above integrable case, the initial momenta were always zero.
Here, in the non--integrable case, one randomly chosen particle is
given a fixed momentum pointing towards the origin. To preserve the
mirror symmetry of the system, the same momentum in opposite direction
is given to the corresponding particle of the second
cloud. 

As follows from Fig.~\ref{fig10}, there is a considerable impact of
the perturbation on the spreading. This means, energy and momentum
must be redistributed to the incoherent, \textit{i.e.},
non--collective, degrees of freedom, because the amplitude of the
collective coordinate decreases the faster the stronger the
perturbation. Nevertheless, we expect the motion of the collective
coordinate to remain largely regular~\cite{hggl} on local time
scales. To further investigate this, we look at the strongest
perturbation with $\lambda=10^{-2}$. We compare simulations with
slightly different initial conditions. In Fig.~\ref{fig12}, are
\begin{figure}
\begin{center}
\includegraphics[scale=1]{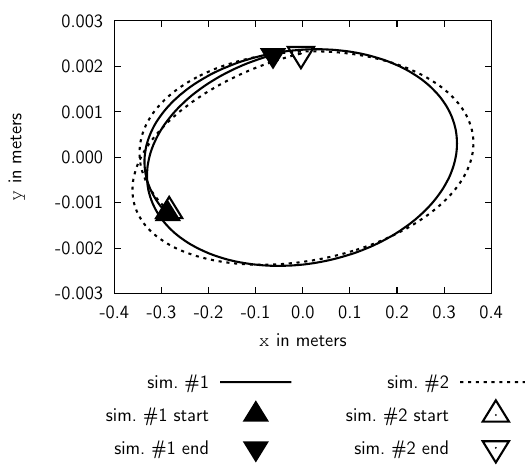}
  \includegraphics[scale=1]{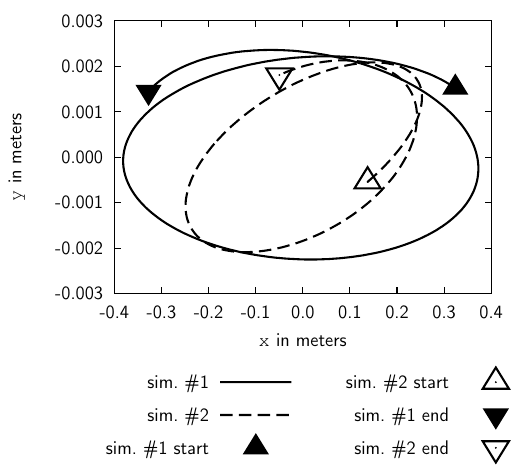}
  \caption{Single-particle trajectories with slightly different
    initial conditions in the two--dimensional position space for
    $\lambda=10^{-2}$ followed for 0.125~s, at shorter times close to
    0~s (top) and at larger times of about 10~s (bottom). We notice a
    factor of 100 between the scales of the $x$ and $y$ direction. The
    beginning and the end of the trajectories are marked with
    triangles, pointing upwards and downwards, respectively. Filled
    and open triangles correspond to the two different trajectories.}
\label{fig13}
\end{center}
\end{figure}
plotted  the collective coordinates for larger times, beyond the time
scale displayed in Fig.~\ref{fig10}. The first deviations are seen in
the region starting at about 6.5~s. The two curves  collapse then
on top of each other again, before they depart once more  at
about 7.5~s. This confirms that the motion of the collective
coordinate remains largely regular. 

To better visualize individual trajectories, we go to two spatial
dimensions by adding the same Hamiltonian that we use for positions
and momenta in $x$ direction also for a new set of positions and
momenta in $y$ direction. The motions in these two direction are thus
uncoupled. As compared to the collective motion, the single--particle
trajectories show a much stronger onset of chaotic motion, as
demonstrated in Fig.~\ref{fig13}. Here, the differences in the initial
conditions can directly be read off from the figure. The two
trajectories are iterated for 10~s. At short times close to 0~s, they
differ only little, but at about 10~s the trajectories are completely
apart. Roughly speaking, the effects due to perturbation accumulate at
later times.

\section{Conclusions}
\label{sec5}

Thermalization in a narrower sense refers to systems of infinitely
many particles.  Among others features, infinite systems have the
advantage of infinite recurrence times, at least in non--integrable
cases.  Nevertheless, the concepts of spreading in closed and of
damping in open  \textit{finite} many--body systems are
intimately related to thermalization. They are of high practical
relevance, since there is a wealth of such systems, particularly
nuclei, atoms, molecules and Bose--Einstein condensates. In parts of
the literature, spreading, damping and thermalization are not clearly
distinguished, and the term thermalization is often used as some kind
of hypernym.

We studied spreading in a closed, selfbound many--body system for
particle numbers large enough to ensure that the recurrence times did
not play a role. Our first result is a clarification: the phenomenon
of spreading is not tied to chaotic motion. We clearly showed that
integrable systems can exhibit spreading.  Thus, chaos is not a
necessary prerequisite. Thermalization, however, always requires
non--integrability. Our second main result is the subtlety of the
effects. Minor modifications in the relative strength parameters or in
the distributions of interaction matrix elements can have a large
impact. The particle number does not necessarily change the time scale
on which spreading occurs.  We explained this by discussing the role
of eigenfrequencies and normal modes. The details of the interactions
are crucial, a normal mode corresponding to the collective coordinate
in question has to exist and has to be isolated from the other
eigenfrequencies and, obviously, it must be excited by the initial
conditions.

Finally, non--integrability, \textit{i.e.}, the onset of chaotic
motion can considerably accelerate spreading. Our third main result is
the different behavior of the collective coordinate and of the single
particles. In accordance with our earlier analytical findings, the
former continues to move in a largely regular fashion, while the
single particles show the onset of chaoticity much stronger.

\section*{Acknowledgements}

We thank Jens H\"ammerling and Sophia Sch\"afer for fruitful
discussions.  We acknowledge support from Deutsche
Forschungsgemeinschaft within the Sonderforschungsbereich Transregio
12.

\end{document}